\documentclass[runningheads]{llncs}
\usepackage{graphicx}
\usepackage{tikz}
\usepackage{amsmath}
\usepackage{graphicx}
\usepackage{graphics}
\usepackage{mathptmx}
\usepackage{amsmath}
\usepackage{amssymb}
\usepackage{outlines}
\usepackage{float}
\usepackage{mathtools}
\usepackage{listings}
\usepackage{algorithm}
\usepackage{pdfrender}
\usepackage{hyperref}

\hypersetup{pdfstartview=FitH,
  pdfpagelayout=SinglePage,
  bookmarksnumbered=true,
  bookmarksopen=true,
  colorlinks=true,
  citecolor=blue,
  linkcolor= blue,
  urlcolor=blue}
\usepackage{wasysym} 
\usepackage{mathrsfs} 

\usepackage{wrapfig}

\usepackage[noend]{algpseudocode}

\usepackage{enumitem}
\setlist[itemize]{topsep=3pt}
\setlist[enumerate]{topsep=3pt}

\usepackage{xspace}
\newcommand{\ie}{{\it i.e.,}\xspace}
\newcommand{\eg}{{\it e.g.,}\xspace}

\newcommand{\parab}[1]{\vspace{0.05in}\noindent\textbf{#1}}

\renewcommand{\figurename}{Fig.}

\usepackage[font=normalsize,labelfont=bf,textfont=it]{caption}
\usepackage{caption,subcaption}
\newsavebox{\largestimage}

\usepackage{hyperref}

\makeatletter
\newcommand{\printfnsymbol}[1]{%
  \textsuperscript{\@fnsymbol{#1}}%
}
\makeatother

\begin{document}

\title{Understanding video streaming algorithms in the wild}
\author{}
\institute{}
\vspace{-0.3in}
%
%

\author{Melissa Licciardello\thanks{equal contribution} \and
Maximilian Gr{\"u}ner \printfnsymbol{1} \and
Ankit Singla}
%
%
\institute{Department of Computer Science, ETH Z{\"u}rich \\ \email{\{melissa.licciardello, mgruener, ankit.singla\}@inf.ethz.ch}}
\maketitle              
\begin{abstract}




While video streaming algorithms are a hot research area, with interesting new approaches proposed every few months, little is known about the behavior of the streaming algorithms deployed across large online streaming platforms that account for a substantial fraction of Internet traffic. We thus study adaptive bitrate streaming algorithms in use at $10$ such video platforms with diverse target audiences. We collect traces of each video player's response to controlled variations in network bandwidth, and examine the algorithmic behavior: how risk averse is an algorithm in terms of target buffer; how long does it takes to reach a stable state after startup; how reactive is it in attempting to match bandwidth versus operating stably; how efficiently does it use the available network bandwidth; etc. We find that deployed algorithms exhibit a wide spectrum of behaviors across these axes, indicating the lack of a consensus one-size-fits-all solution. We also find evidence that most deployed algorithms are tuned towards stable behavior rather than fast adaptation to bandwidth variations, some are tuned towards a visual perception metric rather than a bitrate-based metric, and many leave a surprisingly large amount of the available bandwidth unused.


%

\end{abstract}


\section{Introduction}

Video streaming now forms more than $60\%$ of Internet downstream traffic~\cite{sandvine2019}. Thus, methods of delivering video streams that provide the best user experience despite variability in network conditions are an area of great industry relevance and academic interest. At a coarse level, the problem is to provide a client with the highest possible video resolution, while minimizing pauses in the video stream. There are other factors to consider, of course, such as not switching video resolution often. These considerations are typically rolled into one \underline{q}uality-\underline{o}f-\underline{e}xperience score. Streaming services then use \underline{a}daptive \underline{b}it\underline{r}ate algorithms, which attempt to maximize QoE by dynamically deciding what resolution to fetch video segments at, as network conditions fluctuate.

While high-quality academic work proposing novel ABR is plentiful, the literature is much more limited (\S\ref{sec:related}) in its analysis of widely deployed ABRs, their target QoE metrics, and how they compare to recent research proposals. The goal of this work is precisely to address this gap. Understanding how video platforms serving content to large user populations operate their ABR is crucial to framing future research on this important topic. For instance, we would like to know if there is a consensus across video platforms on how ABR should behave, or whether different target populations, content niches, and metrics of interest, lead to substantially different ABR behavior. We would also like to understand whether ABR research is optimizing for the same metrics as deployed platforms, which are presumably tuned based on operator experience with real users and their measured engagement.

Towards addressing these questions, we present a study of ABR behavior across $10$ video streaming platforms (Table~\ref{tab:providers}) chosen for coverage across their diverse target populations: some of the largest ones in terms of overall market share, some regional ones, and some specialized to particular applications like game streaming (not live, archived). Our methodology is simple: we throttle download bandwidth at the client in a time-variant fashion based on throughput traces used in ABR research, and monitor the behavior of streams from different streaming platforms by analyzing jointly their browser-generated HTTP Archive (HAR) files and properties exposed by the video players themselves.
For robust measurements, we collect data for several videos on each platform, with our analysis herein being based on 6~days of continuous online streaming in total. Our main findings are as follows:

\begin{enumerate}\setlength{\itemsep}{4pt}
    \item Deployed ABRs exhibit a wide spectrum of behaviors in terms of how much buffer they seek to maintain in their stable state, how closely they try to match changing bandwidth vs. operating more smoothly, how they approach stable behavior after stream initialization, and how well they use available network bandwidth. There is thus not a consensus one-size-fits-all approach in wide deployment.
    \item Several deployed ABRs perform better on a QoE metric based on user visual perception rather than just video bitrate. This lends support to the design philosophy of recent ABR work~\cite{cava}, indicating that at least some of the industry is already optimizing towards such metrics rather than bitrate-focused formulations in most prior ABR research.
    \item Most deployed ABRs eschew fast changes in response to bandwidth variations, exhibiting stable behavior. In contrast, research ABRs follow bandwidth changes more closely. It is unclear whether this is due to (a) a mismatch in target metrics used in research and industrial ABR; or (b) industrial ABR being sub-optimal.
    \item Several deployed ABRs leave substantial available bandwidth unused. For instance YouTube uses less than $60\%$ of the network's available bandwidth on average across our test traces. Similar to the above, it is unclear whether this is due to ABR sub-optimality, or a conscious effort to decrease bandwidth costs.
\end{enumerate}


\section{Related Work}
\label{sec:related}

There is a flurry of academic ABR proposals~\cite{oboe,cs2p,bola,mpc,pensieve,squad,piaInfocom,festive,elastic,panda,sabre,cava}, but only limited study of the large number of deployed video streaming platforms catering to varied video types and audiences. 

YouTube itself is relatively well studied, with several analyses of various aspects of its behavior~\cite{candidWithYoutube,anorga2018analysis,wamser2016modeling}, including video encoding, startup behavior, bandwidth variations at fixed quality, a test similar to our reactivity analysis, variation of segment lengths, and redownloads to replace already fetched segments. There is also an end-end analysis of Yahoo's video streaming platform using data from the provider~\cite{yahooAnalysis}.

Several comparisons and analysis of academic ABR algorithms~\cite{yan2019learning,timmerer2016adaptation,stohr2017sweet} have also been published, including within each of the several new proposals mentioned above. In particular, \cite{stohr2017sweet} compares three reference ABR implementations, showing that the configuration of various parameters has a substantial impact on their performance.

Facebook recently published~\cite{icmlFacebook} their test of Pensieve~\cite{pensieve} in their video platform, reporting small improvements (average video quality improvement of $1.6\%$ and average reduction of $0.4\%$ in rebuffers) compared to their deployed approach. 

However, a broader comparative study that examines a large number of diverse, popular streaming platforms has thus far been missing. Note also that unlike ABR comparisons in academic work and head-to-head comparisons of methods in Facebook's study, QoE comparisons across platforms are not necessarily meaningful, given the differences in their content encoding, content type, and audiences. Thus, in contrast to prior work, we define a set of metrics that broadly characterize ABR behavior and compare the observed behavior of a large, diverse set of streaming providers on these metrics. Where relevant, we also contrast the behavior of these deployed ABRs with research proposals. To the best of our knowledge this is the only work to compare a large set of deployed ABRs and discuss how their behavior differs from academic work in this direction.





\section{Methodology}
\label{sec:methods}

\begin{figure*}[t!]
    \centering
    \captionsetup[subfigure]{justification=centering}
    \savebox{\largestimage}{\includegraphics[width=2.3in]{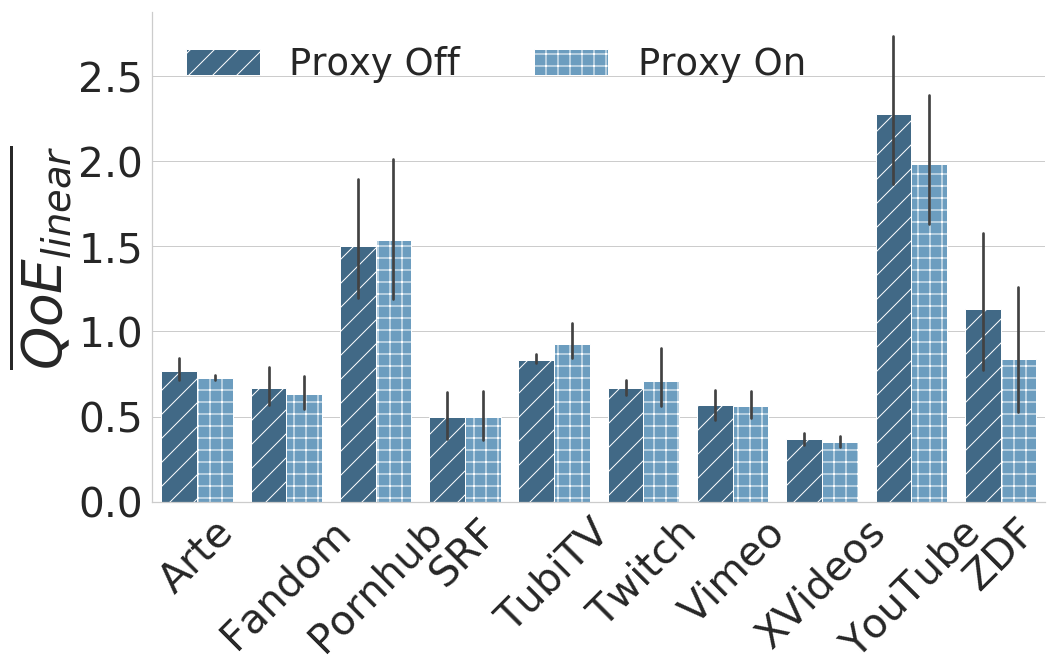}}
    \begin{subfigure}[b]{0.475\textwidth}
        \centering
        \raisebox{\dimexpr.53\ht\largestimage-.47\height}{%
        \includegraphics[width=2.3in]{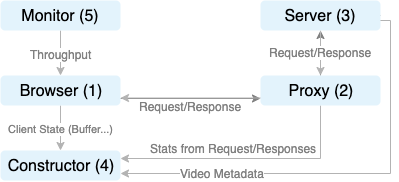}}
        \caption{Experimental setup}
        \label{fig:Setup}
    \end{subfigure}
    \hfill
    \begin{subfigure}[b]{0.475\textwidth}
        \centering
        \usebox{\largestimage}
        \caption{Proxy impact}
        \label{fig:ProxyImpact}
    \end{subfigure}
    \caption{(a) Player behaviour is influenced through bandwidth throttling, and is recorded from multiple sources. (b) The proxy has little impact on player behavior as measured in terms of average linear QoE ($\overline{QoE_{linear}}$); the whiskers are the 95\% confidence interval.}
    \label{fig:SamplingSetup}
\end{figure*}

To understand a target platform's ABR, we must collect traces of its behavior, including the video player's state (in terms of selected video quality and buffer occupancy) across controlled network conditions and different videos. 

\subsection{Experimental setup}

Fig.~\ref{fig:SamplingSetup} shows our architecture for collecting traces about player behaviour. Our Python3 implementation uses the Selenium browser automation framework~\cite{SeleniumWebDriver} to interact with online services. For academic ABR algorithms, trace collection is simpler, and uses offline simulation, as suggested in~\cite{pensieve}.

While playing a video, we throttle the throughput \textbf{(5)} at the client \textbf{(1)} using \texttt{tc} (Traffic control, a Linux tool). The state of the client browser (\eg current buffer occupancy) is captured by the Monitor (\textbf{3}) every $a$ seconds. All requests sent from the client (\textbf{2}) to the server (\textbf{4}) are logged by a local proxy (\textbf{2}). Beyond the final browser state, the proxy allows us to log video player activity such as chunks that are requested but not played. We also obtain metadata about the video from the server (\eg at what bitrate each video quality is encoded). All information gathered from the three sources --- the proxy, the browser and the server --- is aggregated \textbf{(4)}.

Certain players replace chunks previously downloaded at low quality with high quality ones (``redownloading'') in case there is later more bandwidth and no immediate rebuffer risk. Using the proxy's view of requests and responses and the video metadata, we can map every chunk downloaded to a play-range within the video, and use this mapping to identify which chunks / how many bytes were redownloaded.

\parab{How do we add a platform to our measurements? :} Most video platforms (all except YouTube in our set of $10$) use chunk-based streaming. To evaluate such a platform, we use developer tools in Chrome to learn its chunk request format from the video manifest files. This allows us to write code that fetches all chunks for the test videos at all qualities, such that we can use these videos in our offline simulation analysis of the academic ABRs. Having all chunks available also enables calculation of their visual perceived quality (VMAF~\cite{NetflixVMAFBlog}). We also need to map each chunk to its bitrate level and time in the video stream, by understanding how video content is named in the platform (\eg through ``itags'' in YouTube). 

For online experiments through the browser, we need to instrument the platform's video player. We do this by automating the selection of the HTML5 video player element, and having our browser automation framework use this to start the video player and put it in full screen mode. We can then access the current buffer occupancy and current playback time using standard HTML5 attributes.

YouTube does not follow such chunked behavior (as past work has noted~\cite{candidWithYoutube}). It can request arbitrary byte ranges of video from the server. We use an already available tool~\cite{YouTubeDownloader} to download the videos, and then
learn the mapping from the byte ranges to play time from the downloaded videos. 

\subsection{The proxy's impact on measurements}
Some of our measurements (\eg redownloads) use an on-path proxy, so we verify that this does not have a meaningful impact by comparing metrics that can be evaluated without the proxy. For this, we use traces with constant bandwidth $b \in [0.5,0.8,1.2,2.5]$ Mbps, repeating each experiment 5 times for the same video. For our comparison, we calculate QoE using the linear function from Pensieve~\cite{pensieve} with and without the proxy. For every video-network trace combination, we calculate the mean QoE and show the mean across these, together with its 95\% confidence interval with whiskers in Fig.~\ref{fig:SamplingSetup}.

As the results show, for most platforms the proxy has a minimal impact. While there is some impact for YouTube and ZDF, these also show large variations in experiments without the proxy, indicating differing behaviour in very similar conditions in general.

\subsection{Metrics of interest}
\label{subsec:metrics}

Different video platforms serve very different types of content, and target different geographies with varied client connectivity characteristics. It is thus not particularly informative to compare metrics like QoE across platforms. For instance, given the different bitrate encodings for different types of content, QoE metrics using bitrate are not comparable across platforms. We thus focus on comparisons in terms of the following behavioral and algorithm design aspects.

\parab{Initialization behavior:} We quantify how much \textit{wait time} a video platform typically incurs for streams to start playback, and how much \textit{buffer} (in seconds of playback) it builds before starting. 
We use traces with a fixed bandwidth of $3$~Mbps until player's HTML5 interactions are available, thus always downloading items like the player itself at a fixed bandwidth. After this, we throttle using only the high-bandwidth traces from the Oboe~\cite{oboe} data set, which have a mean throughput of $2.7$~Mbps.
We start timing from when the first chunk starts downloading (per the HAR files; the player HTML5 interactions may become available earlier or later).

\parab{Convergence:} During startup, an ABR may have little information about the client's network conditions. How do different ABRs approach stable behavior starting from this lack of information? Stablility in this sense refers to fewer bitrate switches. Thus, to assess convergence characteristics, we quantify the bitrate changes (in Mbps per second) across playback, \ie a single switch from $3$~Mbps to $4$~Mbps bitrate over a total playback of $5$-seconds amounts to $0.2$ Mbps/sec on this metric.


\parab{Risk-tolerance:} ABRs can hedge against rebuffer events by building a larger buffer, thus insulating them from bandwidth drops. Thus, how much \textit{buffer} (in seconds of video) an ABR builds during its stable operation is indicative of its risk tolerance.

\parab{Reactivity:} ABRs must react to changes in network bandwidth. However, reacting too quickly to bandwidth changes can result in frequent switching of video quality, and cause unstable behavior when network capacity is highly variable. To quantify reactivity of an ABR, we use synthetic traces with just one bandwidth change after convergence, and measure the evolution of \textit{bitrate difference} in the video playback after the change over time (with the number of following chunk downloads used as a proxy for time).

\parab{Bandwidth usage:} ABR must necessarily make conservative decisions on video quality: future network bandwidth is uncertain, so fetching chunks at precisely the estimated network bandwidth would (a) not allow building up a playback buffer even if the estimate were accurate; and (b) cause rebuffers when bandwidth is overestimated. Thus, ABR can only use some fraction of the available bandwidth. We quantify this behavior in terms of the fraction of \textit{bytes played to optimally downloadable}, with ``optimally downloadable'' reflecting the minimum of (\textit{a posteriori} known) network capacity and the bytes needed for highest quality streaming. 

For better bandwidth use and to improve QoE, some ABRs are known to redownload and replace already downloaded chunks in the buffer with higher quality chunks. We quantify this as the fraction of \textit{bytes played to bytes downloaded}. Fractions $<$$1$ reflect some chunks not being played due to their replacement with higher quality chunks.

\parab{QoE goal:} Academic ABR work has largely used a QoE metric that linearly combines a reward for high bitrate with penalties for rebuffers and quality switches~\cite{mpc,pensieve}. More recent work has suggested formulations of QoE that reward perceptual video quality rather than just bitrate~\cite{cava}. One such metric of perceptual quality, VMAF~\cite{NetflixVMAFBlog}, combines several traditional indicators of video quality. While it is difficult, if not impossible, to determine what precise metric each platform's ABR optimizes for, we can evaluate coarsely whether this optimization is geared towards bitrate or VMAF-like metrics by examining what video chunks an ABR tries to fetch at high quality: do chunks with higher VMAF get fetched at a higher quality level? 
To assess this, we sort chunks by VMAF (computed using~\cite{NetflixVMAFBlog}) and quantify for the top $n\%$ of chunks, their (average) playback quality level compared to the (average) quality level of all chunks, $\overline{Q_{top-n\%}} - \overline{Q_{all}}$. A large difference implies a preference for high-VMAF chunks.

\subsection{Measurement coverage} We evaluate multiple videos on each of $10$ platforms across a large set of network traces.

\begin{table*}[b]
\centering
\setlength{\tabcolsep}{5pt}
\renewcommand*{\arraystretch}{1.3}
\begin{tabular}{lll}
\textbf{Provider} & \textbf{Description} & \textbf{\# Resolutions offered} \\
\hline
\textbf{Arte} & French-German, cultural& $4.0 \pm 0.0$ \\
  \textbf{Fandom}& Gaming, pop-culture & $5.0 \pm 0.0$ \\
  \textbf{SRF} &Swiss Public Service, local and international content & $5.7 \pm 0.48$ \\
 \textbf{TubiTV} & Movies and series of all genres& $3.0 \pm 0.0$ \\
 \textbf{Twitch} & Live and VoD streaming service, gaming& $5.9 \pm 0.32$ \\
 \textbf{Vimeo} & Artistic content~\cite{WhatIsVimeo}  & $4.2 \pm 0.92$ \\
 \textbf{YouTube} & Broad coverage & $6.5 \pm 1.08$ \\
 \textbf{ZDF} & German Public Service, local and international content & $5.3 \pm 0.48$ \\
 \textbf{Pornhub} &  Pornographic video sharing website & $4.0 \pm 0.0$ \\
 \textbf{XVideos} &  Pornographic video sharing website & $4.4 \pm 0.52$ \\
\hline
\end{tabular}
\vspace{8pt}
\caption{We test a diverse set of large video platforms.}
\label{tab:providers}
\vspace{-12pt}
\end{table*}

\parab{Target platforms:} Table~\ref{tab:providers} lists the platforms for which we have currently implemented support in our measurement and analysis pipeline. While by no means exhaustive, these were chosen to cover a range of content types and a few different geographies. Note that Netflix was excluded because their terms of service prohibit automated experiments~\cite{netflixTerms}. For Twitch, which offers both live streams and video-on-demand of archived live streams, we only study the latter, as live streaming is a substantially different problem, and a poor fit with the rest of our chosen platforms.

Different platforms encode content at varied resolutions and number of resolutions, ranging from just $3$ quality levels for TubiTV to $6.5$ on YouTube (on average across our test videos; YouTube has different numbers of resolutions on different videos.)

When comparing the behavior of deployed ABRs with academic ones, we test the latter in the offline environment made available by the Pensieve authors~\cite{pensieve}. For each tested video on each platform, we pre-download all its chunks at all available qualities. We then simulate playback using the same network traces up until the same point offline for academic ABRs as we do for the deployed ones. We primarily rely on Robust MPC~\cite{mpc} (referred to throughout as MPC) as a stand-in for a recent, high-quality academic ABR approach. While even newer proposals are available, they either use data-dependent learning techniques~\cite{pensieve,oboe} that are unnecessary for our purpose of gaining intuition, or do not have available, easy-to-use code.

\parab{Videos:} The type of content can have substantial bearing on streaming performance, \eg videos with highly variable encoding can be challenging for ABR. We thus used a set of $10$ videos on each platform. Where a popularity measure was available, we used the most popular videos; otherwise, we handpicked a sample of different types of videos. Videos from each platform are encoded in broadly similar bitrate ranges, with most differences lying at higher qualities, \eg some content being available in $4$K.

It would, of course, be attractive to upload the same video content to several platforms (at least ones that host user-generated content) to remove the impact of videos in the cross-platform comparisons. However, different platforms use their own encoding pipelines, making it unclear whether this approach has much advantage over ours, using just popular videos across platforms.


\parab{Network traces:} Our experiments use synthetic and real-world traces from 3 datasets in past work~\cite{oboe,NorwayTraces,FCCTraces}. Unfortunately, a full cross-product of platform-video-trace would be prohibitively expensive --- the FCC traces~\cite{FCCTraces} alone would require $4$ years of streaming time. To sidestep this, we rank traces by their throughput variability and pick traces with the highest and lowest variability together with some randomly sampled ones. 

Our final network trace collection consists of the 5 least stable, 5 most stable, and 5 random traces from the Belgium trace collection~\cite{BelgiumTraces}, and 10 in each of those categories from the Norway~\cite{NorwayTraces}, the Oboe~\cite{oboe} and the FCC datasets\footnote{Specifically, the stable collection from September 2017~\cite{FCCTraces}.}. We also use 15 constant bandwidth traces covering the range from $0.3$ to $15$ Mbps uniformly. Lastly we add 10 step traces: after 60 seconds of streaming we suddenly increase/drop the bandwidth from/to 1 Mbps to/from 5 values covering the space from $1.5$ to $10$ Mbps uniformly. 

In total, we use $130$ traces with throughput (average over time for each trace) ranging from $0.09$ to $41.43$~Mbps, with an average of $6.13$~Mbps across traces. Note that we make no claim of our set of traces being representative; rather our goal is to test a \textit{variety} of traces to obtain insight into various ABR behaviors.
If a trace does not cover the whole experiment we loop over it.

For quantifying reactivity, we only use the synthetic traces mentioned above, with a single upward step change in bandwidth. 
For quantifying startup delay, we use traces with a bandwidth of around 3~Mbps as noted in \S\ref{subsec:metrics}.

\parab{Ethics:} We are careful to not generate excessive traffic or large bursts to any platform, measuring at any time, only one stream per service, typically at a low throttled rate. 
\begin{figure*}[t]
    \centering
    \begin{subfigure}[t]{0.475\textwidth}
    \captionsetup[subfigure]{justification=centering}
        \centering
        \includegraphics[width=2.3in]{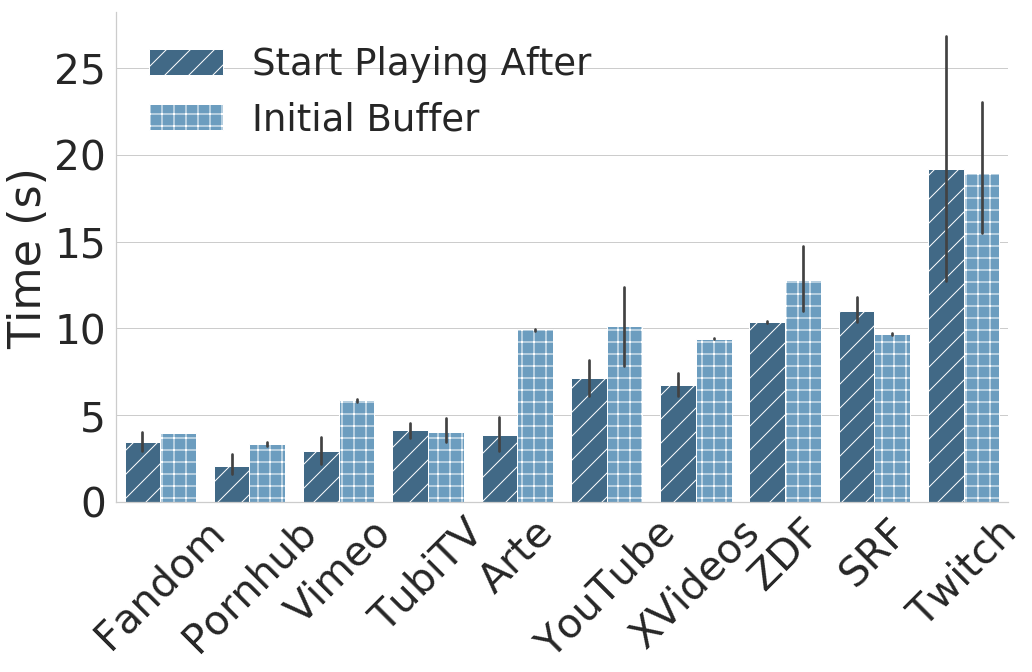}
        \caption{Initialization behavior}
         \label{fig:initialization}
    \end{subfigure}
    \hfill
    \begin{subfigure}[t]{0.475\textwidth}
        \centering
        \raisebox{0.15in}{
        \includegraphics[width=2.3in]{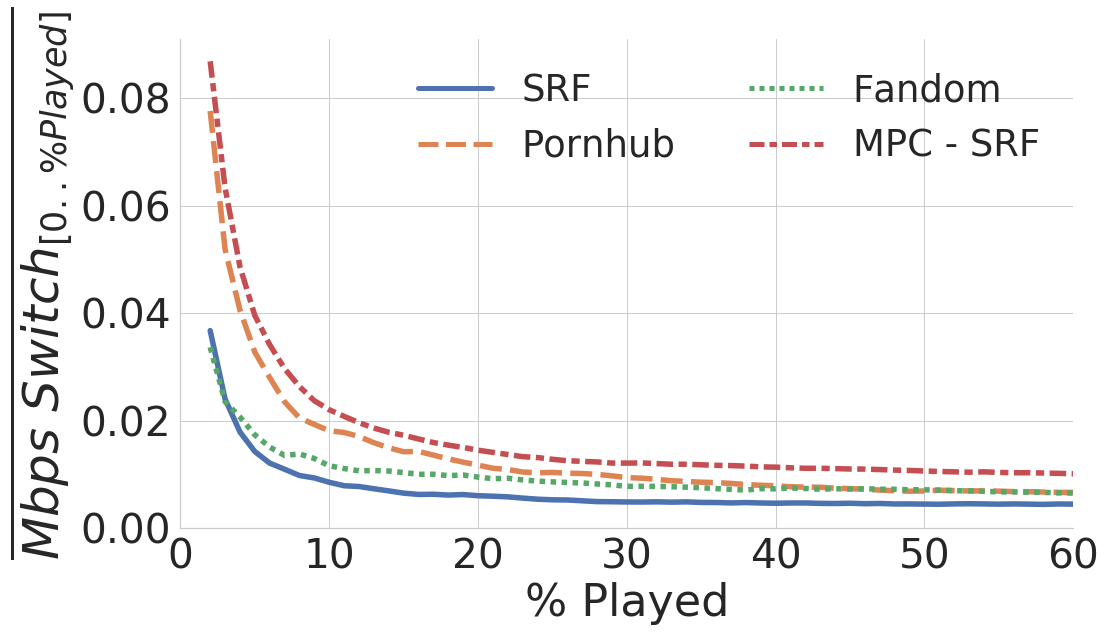}}
        \caption{Convergence}
        \label{fig:convergence}
    \end{subfigure}
    \caption{(a) Initialization behavior: most providers start playback after one chunk is downloaded. (b) Convergence is measured in terms of changes in bitrate switching, \ie the (absolute) sum of bitrate differentials across all switches from the start until a point in playback, divided by the thus-far playback duration. As expected, switching is more frequent during startup, but the degree of switching varies across providers both in startup and later.}
    \label{fig:affinityTolerance}
\end{figure*}

\section{Measurement results} 
\label{sec:results_feature}



Overall, we find diverse behavior on each of our tested metrics across the measured platforms. We attempt to include results across all platforms where possible, but for certain plots, for sake of clarity, we choose a subset of platforms that exhibits a range of interesting behaviors.

\parab{Initialization behavior, Fig.~\ref{fig:initialization}:} We find that most platforms' ABR simply waits for one chunk download to finish before beginning playback. This is reflected in the buffer occupancy at playback. Some players like ZDF and SRF use a larger chunk size ($10$ seconds), which is why they pre-load more seconds of buffer. 

As one might expect, building a larger buffer before playback starts generally incurs a higher start time. Twitch stands out in this regard, as it downloads nearly $20$~seconds of buffer before start. Some players, whilst downloading the same number of buffer seconds as others, do so at much higher resolution -- \eg SRF downloads its first $10$ seconds with $6\times$ as many pixels as Arte. This is reflected in the disparity between their start times, despite both populating the buffer with $10$~seconds of playback. More broadly, all such ``discrepancies'' are difficult to explain because startup is hard to untangle from other network activity, \eg some players already start downloading video chunks while the player itself is still downloading, thus complicating our notion of timing. (We start timing from the point the first chunk starts downloading.  For most platforms, this provides a leveling standard that excludes variation from other downloads on their Web interface. It also helps reduce latency impacts that are mainly infrastructure driven, as well as effects of our browser automation framework.)

\parab{Convergence, Fig.~\ref{fig:convergence}:} As one might expect, during startup and early into playback, every player attempts to find a stable streaming state. This results in a large amount of bitrate switches early in playback followed by much smoother behavior with more limited switching. Nevertheless, there are large differences across players, \eg Pornhub switches more than twice as much as Fandom and SRF in the beginning. In stable state, Fandom switches substantially more than SRF. We also evaluated the academic (Robust) MPC algorithm~\cite{mpc} on the same network traces and over the SRF videos. 
The MPC algorithm would use more than twice as much switching both in startup and later, compared to SRF's deployed ABR. It is unclear to us whether SRF's ABR is sub-optimal, or whether their deployment experience indicates stability has a higher importance than reflected in the default linear QoE model used in MPC.

For clarity, we only picked a few platforms as exemplars of behavior towards convergence instead of including all $10$ tested platforms. The behavior is broadly similar with more switching early on, but the precise stabilization differs across platforms. 

\begin{wrapfigure}{r}{2.3in}
        \vspace{-28pt}
        \setlength\intextsep{0pt}
        \begin{center}
           \includegraphics[width=2.3in,clip]{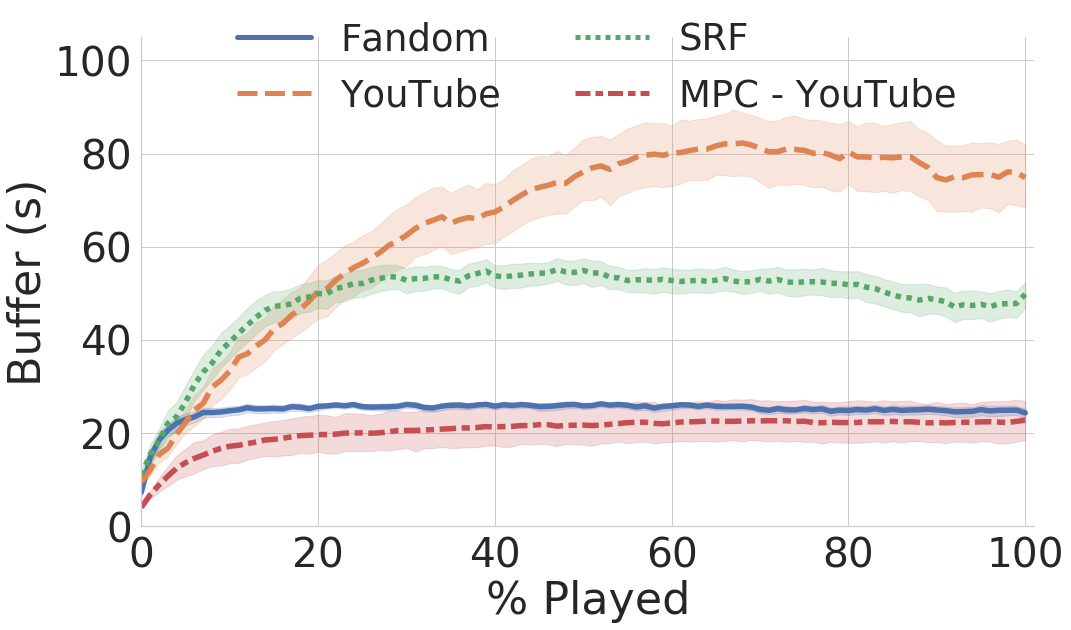}
           \caption{Risk-tolerance: YouTube operates with nearly $4\times$ the buffer for Fandom. The shaded regions show the $95\%$ confidence interval around the mean.}
           \label{fig:riskTolerance}
           \vspace{-26pt}
        \end{center}
\end{wrapfigure}

\parab{Risk-tolerance, Fig.~\ref{fig:riskTolerance}:} We observe widely different buffering behavior across the players we tested. Of course, every player uses early playback to download lower quality chunks and accumulate buffer, but some, like YouTube, settle towards as much as $80$~seconds of buffer, while others like Fandom operate with a much smaller buffer of around $20$~seconds. 
Testing MPC's algorithm on the same traces across the YouTube videos reveals that it falls towards the lower end, stabilizing at $20$~seconds of buffer. 

Note that for approaches that allow redownloads (including YouTube), larger buffers are a reasonable choice: any chunks that were downloaded at low quality can later be replaced. This is likely to be a more robust strategy in the face of high bandwidth variability. However, for approaches that do not use redownloads, a larger buffer implies that all its content must be played out at whatever quality it was downloaded at, thus limiting the possibilities to benefit from opportunistic behavior if bandwidth later improves. Thus operating with a smaller buffer of higher-quality chunks may be preferable to filling it with lower-quality chunks. In the absence of redownloads, there is thus a tradeoff: a larger buffer provides greater insurance against bandwidth drops, but reduces playback quality. At the same time, redownloads are themselves a compromise: \emph{if} better bitrate decisions could be made to begin with, redownloads amount to inefficient bandwidth use.

\begin{figure}[t]
\includegraphics[width=\textwidth]{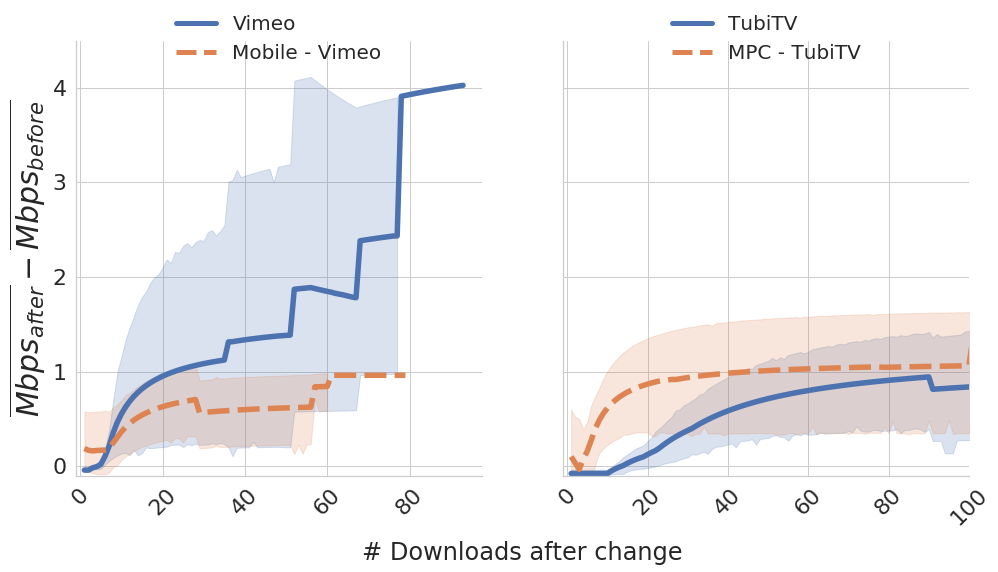}
 \caption{We measure reactivity in terms of bitrate evolution after a bandwidth increase, \ie difference in average playback bitrate after and before the bandwidth change over time (in terms of chunk downloads). The plots show the reactivity differences between: (left) mobile and desktop versions of Vimeo; and (right) TubiTV and MPC.}
\label{fig:switching_factor}
\end{figure}
\parab{Reactivity, Fig.~\ref{fig:switching_factor}:} We find that most deployed ABRs are cautious in reacting to bandwidth changes. This is best illustrated through comparisons between deployed and academic ABRs. Fig.~\ref{fig:switching_factor}(right) shows such a comparison between TubiTV and MPC evaluated on the same traces and videos. After the bandwidth increases (at $x$-axis=0 in the plot), TubiTV waits for tens of chunk downloads before it substantially ramps up bitrate. In contrast, MPC starts switching to higher bitrates within a few chunk downloads. (The large variations around the average arise from the varied sizes of the step-increases in the used network traces and variations in the tested videos.) 


While we have not yet been able to evaluate a large number of mobile ABR implementations (see \S\ref{sec:limitations}), we were able to experiment with Vimeo's mobile and desktop versions, shown in Fig.~\ref{fig:switching_factor}(left). They exhibit similar ramp-up behavior in terms of how many downloads it takes before Vimeo reacts, but show very different degrees of bitrate change. The desktop version increases bitrate in several steps after the bandwidth increase, while the mobile one settles at a modest increase. This is along expected lines, as the mobile player, targeting the smaller screen, often does not use the higher-quality content at all.

A comparison between TubiTV and Vimeo (desktop) across the two plots is also interesting: Vimeo ramps up faster than TubiTV. (MPC ramps us even faster on the Vimeo videos.) One potential reason is the difference in encoding --- TubiTV serves each video in only 3 resolutions, compared to Vimeo's 4-5. This implies that over the same network traces, TubiTV must necessarily see a larger change in bandwidth to be able to jump from one bitrate to the next, given its larger differential in bitrate levels.

\begin{figure*}[t!]
    \centering
    \captionsetup[subfigure]{justification=centering}
    \begin{subfigure}[t]{0.475\textwidth}
        \centering
        \includegraphics[width=2.3in]{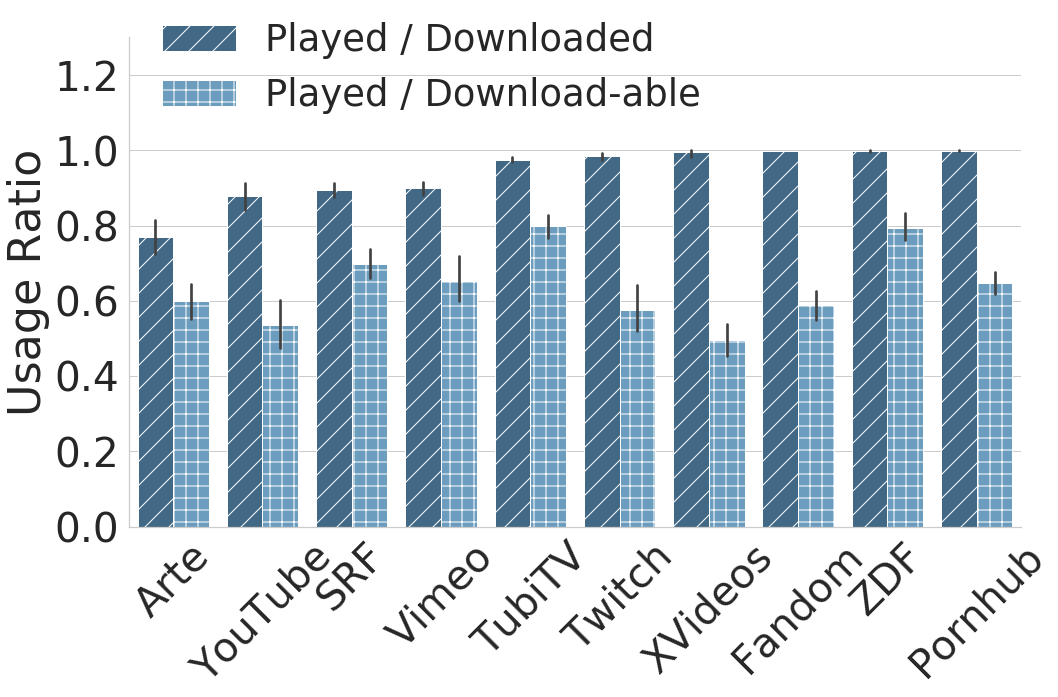}
        \caption{Bandwidth usage}
         \label{fig:datawastage}
    \end{subfigure}
    \hfill
    \begin{subfigure}[t]{0.475\textwidth}
        \centering
        \raisebox{0.15in}{
        \includegraphics[width=2.3in]{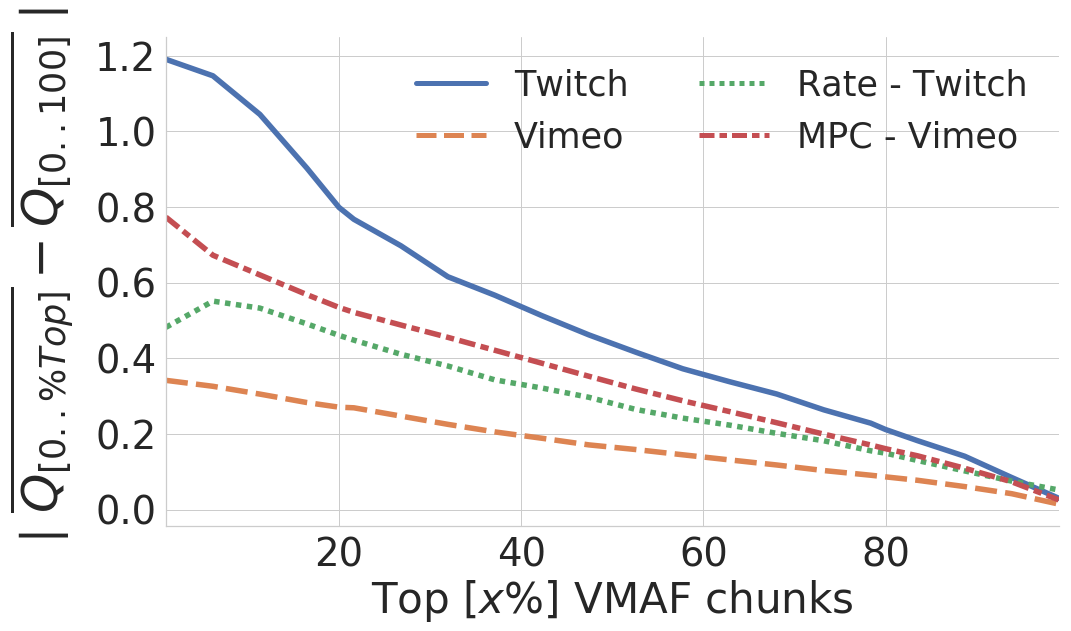}}
        \caption{QoE goal}
        \label{fig:qoe}
    \end{subfigure}
    \caption{(a) Bandwidth usage: many players use surprisingly little of the available network bandwidth (Played / Download-able), \eg XVideos uses only $50\%$ of it; and some players, like Arte, spend a large fraction of their used bandwidth on redownloads. (b) QoE goal: we measure how much a player prefers high-VMAF chunks by quantifying the average quality-level difference between all chunks and only the top-$x\%$ of chunks by VMAF (\ie $\overline{Q_{[0\ldots \%Top]}}$).
    Some players, like Twitch, show a large preference for high-VMAF chunks.}
    \label{fig:wastageQoE}
\end{figure*}


\parab{Bandwidth usage, Fig.~\ref{fig:datawastage}:} Different platforms use bandwidth very differently. Arte discards a surprisingly large $23\%$ of its downloaded bytes in its efforts to replace already downloaded low-quality chunks with high-quality ones. Some platforms, including YouTube, SRF, and Vimeo, show milder redownload behavior, while several others, including XVideos, Fanrom, Pornhub, and ZDF, do not use redownloads at all.

In terms of efficiency, ZDF and TubiTV are able to use $80\%$ of the network's available bytes for fetching (actually played) video chunks, while all other players use the network much less effectively. While the uncertainty in future bandwidth and the desire to maintain stable streaming without many quality switches \emph{necessitates} some bandwidth inefficiencies, we were surprised by how large these inefficiencies are. In particular, XVideos, YouTube, Twitch, and Fandom all use less than $60\%$ of the network's available capacity on average across our trace-video pairs\footnote{Note that these inefficiencies cannot be blamed on transport / TCP alone, as on the same traces, other players are able to use $80\%$ of the available capacity. We also carefully account for non-video data to ensure we are not simply ignoring non-chunk data in these calculations. For instance, audio data is separately delivered for Vimeo and YouTube, but is accounted for appropriately in our bandwidth use analysis.}. This low usage is particularly surprising for YouTube, which uses several strategies --- variable chunk lengths (as opposed to fixed-size chunks in other providers), larger number of available video resolutions, and redownloads --- that allow finer-grained decision making, and thus should support more effective bandwidth use. Given these advanced features in their ABR design, it is more likely that their optimization goals differ from academic ABR work than their algorithm simply being poorly designed. While we cannot concretely ascertain their optimization objectives, one could speculate that given the large global demands YouTube faces while operating (largely) as a free, ad-based service, a profit maximizing strategy may comprise providing good-enough QoE with a limited expense on downstream bandwidth.

\parab{QoE goal, Fig.~\ref{fig:qoe}:} We find that some providers fetch high-VMAF chunks at higher quality than the average chunk. In particular, Twitch fetches the chunks in the top $20$\textsuperscript{th} percentile by VMAF at a mean quality level $0.79$ higher than an average chunk. If instead of Twitch's ABR, we used a VMAF-unaware, simple, rate-based ABR\footnote{This ABR estimates throughput, $T$, as the mean of the last $5$ throughput measurements. For its next download, it then picks the highest quality level with a bitrate $\leq T$. It thus downloads the largest chunk for which the estimated download time does not exceed the playback time.} that uses an estimate of throughput to decide on video quality, this difference in quality level between high-VMAF and the average chunk would reduce to $0.46$. 

Note that given the correlation between higher quality and higher VMAF, high-VMAF chunks are overall more likely to be fetched at high quality; what is interesting is the degree to which different players prefer them. Vimeo, for instance, shows a much smaller difference of $0.27$ between the quality level of chunks in the top $20$\textsuperscript{th} percentile and an average chunk. If MPC's ABR were used to fetch chunks from Vimeo, this difference increases to $0.534$, because MPC is willing to make more quality switches than Vimeo.

Our results thus indicate diversity in optimization objectives in terms of bandwidth use and QoE targets across deployed video platforms. It is at least plausible that academic ABRs produce different behavior over the same traces not because they are much more efficient, but rather the optimization considerations are different. While algorithms like MPC are flexible enough to be used for a variety of optimization objectives, it is unclear how performance would compare across a suitably modified MPC (or other state-of-the-art ABR) when evaluated on operator objectives.

\section{Limitations and future work}
\label{sec:limitations}

Our first broad examination of a diverse set of widely deployed ABRs reveals several interesting insights about their behavior, but also raises several questions we have not yet addressed:

\begin{enumerate}\setlength{\itemsep}{3pt}
    \item Does ABR behavior for the same platform vary by geography and client network? Such customization is plausible --- there are likely large differences in network characteristics that a provider could use in heuristics, especially for startup behavior, where little else may be known about the client's network bandwidth and its stability. However, addressing this question would require running bandwidth-expensive experiments from a large set of globally distributed vantage points.
    \item How big are the differences between mobile and desktop versions of ABR across platforms? Unfortunately, while the browser provides several universal abstractions through which to perform monitoring on the desktop, most platforms use their own mobile apps, greatly increasing the per-platform effort for analysis.
    \item If we assume that the largest providers like YouTube and Twitch are optimizing ABR well, based on their experience with large populations of users, can we infer what their optimization objective is? While there are hints in our work that these providers are not necessarily optimizing for the same objective as academic ABR, we are not yet able to make more concrete assertions of this type.
    \item Does latency have a substantial impact on ABR? ABR is largely a bandwidth-dependent application, but startup behavior could potentially be tied to latency as well. We have thus far not evaluated latency-dependence.
\end{enumerate}



\section{Conclusion}

We conduct a broad comparison of adaptive bitrate video streaming algorithms deployed in the wild across $10$ large video platforms offering varied content targeted at different audiences. We find large differences in player behavior, with a wide spectrum of choices instantiated  across virtually all metrics we examined. For instance, our results show that: (a) some deployed ABRs are conscious of perceptual quality metrics compared to others focused on bitrate; (b) no deployed ABRs follow available bandwidth as closely as research ABRs; and (c) several ABRs leave a large fraction of available network capacity unused. Whether this diversity of design choices and behaviors stems from careful tailoring towards different use cases and optimization objectives, or is merely a natural consequence of sub-optimal, independent design is at present unclear. But if large, otherwise extremely well-engineered platforms like YouTube differ so substantially from state-of-the-art research ABRs, then it is at least plausible that ABR research is more narrowly focused than desirable.


\bibliographystyle{splncs04} 
\bibliography{main} 
\newpage

\end{document}